# On *Quanta of Information* and Electromagnetic Fields–

# The Information-theoretic Origin and Structure of Quantum Fields

Masroor H. S. Bukhari
Department of Physics, Jazan University, Jazan 45142, Saudi Arabia
mbukhari@jazanu.edu.sa
ORCID: https:// https://orcid.org/0000-0003-3604-3152

## Abstract

A preliminary model of meaningful quanta of information arising from quantum vacuum, and giving rise to the elementary quantum fields, is presented. We attempt to establish an equivalence of information and physical action, with the quanta (or packets of words) of information as a source of quantum fields (with a possible role in governing the interactions among those), similar to the quantum fields being the source of classical information. The information and quantum fields emerge in an entangled form out of the quantum vacuum, with their common underlying basis bearing intrinsically both a well-defined deterministic (geometric) and probabilistic (stochastic) structure. The description of equivalent energy of a quantum of the electromagnetic field, in terms of its information content, as well as an expression of the signal power, is provided. Various aspects of this problem and its possible implications and repercussions on our understanding of elementary quantum fields and their interactions are discussed. It is suggested that information, in a deterministic form, as well as chaotic processes within the quantum vacuum give rise to a possible *"condensation"* of information into quantum fields.

## Introduction

The nature and dynamics of information, most particularly quantum information, are an extremely important and sought-after area of contemporary studies. Followed by consorted efforts within the regimes of string theory and various models of quantum gravity during the last many decades, investigating the physical basis of all the matter that exists in our universe in various forms, and possibly of the origins of the universe and space-time itself, there has been a recent paradigm shift in the form of information physics. Understanding the structure and form of information and its connections with topology and other deep symmetries in nature which give rise to strings and/or fundamental fields of matter may be one of the most probable routes to understanding the physical reality of everything.

It has been suggested that everything may have information as its basis [1], and consequently all the matter and radiation may have been created as a result of information. This conjecture results into equivalence of information and energy, in analogy to the established equivalence between the latter and inertial mass. There are important consequences of this idea on our understanding of the physical universe, matter and quantum fields. The most profound implication is that the universe has entropic origins, with entanglement of space-time with information. Moreover, as a consequence, the universe may also be pixelated, including all the forms of matter and energy, and possibly the missing mass and energy in the universe (the so-called *dark sector*) as well, with base of everything in the form of a (possibly binary) code. There may be raised a number of logical arguments against this claim. The most likely possibility is that information and energy may both exist independently and in conjunction with each other (under some fundamental physical interaction) to create matter and radiation, instead of information giving rise to them *ab initio* or independently. However, a tenuous support comes from the black hole thermodynamics, where entropy (information) is recovered when no other quantity is retrievable.

Mass generation is often attributed to a Higgs field [2] which endows mass to all the fields (massless in nature, to begin with), which become massive after interaction with the former, following spontaneous symmetry breaking. However, the Higgs field was itself created (most likely during the thermalization epoch of the early moments of the universe after its creation), along with all the other fields in the "*Standard Model*" (and possibly those beyond the SM as well). Thus, there has to be a more fundamental mechanism of mass and energy generation which existed at the time of initial creation of our universe (and of all physical universes in general). It is quite plausible that information carried within or coupled with vacuum fluctuations may have been the source or might have played an important role in primordial mass and energy generation during the *Planck era*, before the onset of inflation and thermalization epochs. It is well-established that the quantum vacuum is of an electromagnetic form in nature and contains all the possible electromagnetic modes $k$ (with $\omega \neq 0$ where $\omega$ stands for the angular frequency) as well as ground state fluctuations which give rise to transient virtual photons of all the possible modes which can end up in generating pairs of particles and anti-particles out of the vacuum (vacuum polarization).

It is attempted here to present a fundamental model in which a form of legitimate information couples to an electromagnetic field to give rise to the physical action, and the interaction of the two may possibly become the underlying basis of quantum fields creation, and thus information transcends into energy in conjunction with quantum fluctuations.

It is suggested in the end that there may be *"quanta of information"* (initially formed within the fabric of quantum vacuum) similar to the quanta of fields which give rise to and interact with the quantum fields (or strings within the String theoretical framework, whatever may be the fundamental vibration topology of matter and radiation). It is cited as a possibility in the end that the existence of the information content which translates to physical action can be detected within a volume of quantum vacuum in the form of well-defined glitches above the noise floor with sensitive low-temperature (T<<) experiments [3], as are there to detect cold dark matter particles and other fields beyond the SM.

*The Model:*

Based upon underlying symmetries in nature, out of some symmetry Generators $\vartheta$, it is suggested here that the information Z exists spontaneously or in a stimulated form within the quantum vacuum, and is intrinsically entangled with quantum fluctuations *in vacuo*, such that a physical action may rise and result into the generation of an electromagnetic scalar field (a photon) of four-momentum $p$ (physical energy $E$ and $\boldsymbol{p}$), with zero inertial mass and no electrical charge. The photon, if having sufficient energy ($E_0 \geq 2m_0 c^2$), can result into the production of a pair of leptons or quarks, under the electromagnetic interaction. The information has both probabilistic and deterministic components characterized by a stochastic (Markovian) and a binary weight term, respectively.

The preliminary model we consider here is based upon assumptions that the information is binary in nature (i.e. having only two possibilities, since there is either presence of information or nothing within a volume of space, and conversely an existence of an excitation or nothing) and is carried forward in space entangled with the quanta of electromagnetic fields, in the form of an interaction governed by both frequency of the field modes as well as probability for the interaction to take place, which can be expressed in the form of a joint probability.

A "*word*" of meaningful *information* spontaneously, or in stimulated form, arising out of the (quantum) vacuum, which may give rise to a physical action, can be written in its simplest form as a probability-weighted summation of its information content:

$$Z = \sum_i^n \Pi(n) \zeta_n \quad \underline{\quad\quad\quad\quad} \quad (1)$$

Here, the $\zeta$ depicts a "*quantum of information*" and $\Pi$ is the probability weight for the particular information quantum to be present in or arising out of the vacuum.

The fundamental information quantum (assuming that it is in the form of a binary, ie $2^n$, form) is written as:

$$\zeta_n = \frac{2^n \ln 2}{2\pi} \quad \underline{\quad\quad\quad\quad} \quad (2)$$

With the minimum *quantum of information*, or an *elementary information quantum bit (qubit)*, defined as[§]:

$$\zeta_0 = \frac{\ln 2}{2\pi} \quad \underline{\quad\quad\quad\quad} \quad (3)$$

Such that it corresponds to the minimum quantum of energy (i.e. the zero-point mode):

$$E_0 = \frac{1}{2\pi} h\omega \quad \underline{\qquad} (4)$$

where $\omega$ is the angular frequency of a mode *k* of electromagnetic field corresponding to the nth word of information. Similar to a quantum of electromagnetic energy, $E_0$, or a quantum of thermal conductance, $g_0$, the "$\zeta_0$" is the non-vanishing pedestal of information, an elementary information quantum.

Then, in an d-dimensional Hilbert space $\mathfrak{H}$, the information, $Z(\omega)$, in an entangled form with the quanta of electromagnetic fields, and translated into physical action, can be expressed in a generalized form as:

$$Z(\omega) = hT^{-1} S(k,n) \lambda_{k,n} \omega_k \Pi(n) \zeta_n \quad \underline{\qquad} (5)$$

Where T is temperature, $\lambda$ is the coupling between the information quantum and the mode (*k*) of the electromagnetic field, interacting with the former. The structure and value of this coupling shall be determined later. $\Pi(n)$ is the probability of finding the *nth* word. $S(k,n)$ is the joint probability density matrix of information and electromagnetic modes quanta (it becomes the entanglement entropy of the nth information word entangled with the *kth* mode of the field, analogous to a bi-dimensional *Shannon entropy*[§], but not equal to and a superset of it). For a detailed treatment of the stochastic contributions to this expression, in terms of *Markovian* and involved quantum stochastic density matrix formulation, any of the relevant texts may be seen, such as [6]. A suitable *Lévy process*, or its specialized continuous form, such as a *Weiner process* [7], can be used to capture the stochastic content of the interaction and its dynamics.

The result is a finite semi-classical information density, obtained from an entanglement of information with the electromagnetic fields. The information, as well as its entangled form with the fields quanta, are both not observable, the only observable quantities arising out of the interaction are the energy eigenvalues (corresponding to appropriate density matrices). A quantum measurement would correspond to measuring the density matrix which encodes the underlying entangled (*mixed*) state, and in that case, the resulting state entropy shall have a *Von Neumann* form [8], carrying the information content in the appropriate density matrix form, as:

$$S(\chi) = -tr[\rho(\chi) ln\rho(\chi)] \quad \underline{\qquad} (7)$$

_____

[§] The Shannon entropy [4, 5], *bi-dimensional* in this case, is defined here as:

$$S(x,y) = -\sum_x \sum_y \Pi(x,y) ln[\Pi(x,y)] \quad \underline{\qquad} (6)$$

Where $\Pi(x,y)$ is the probability of two entangled variables x and y.

Here $\chi$ is a suitable time-dependent stochastic process and $\rho$ the density matrix which encodes the probability of finding a particular information content ($\Pi(n)$), or the *nth* word, entangled with a mode $k$, in a stochastic manner. The density matrix $\rho$ carries within it the bipartite entanglement as quantified by the *Von Neumann* entropy function, and thus this function is the most appropriate measure of both the degree of entanglement and of the information content. An average value of the entropy, its probability distribution and time evolution can thus be found and would give valuable insights into the interaction itself and its dynamics.

It is straightforward then, to obtain the energy of resulting fields or matter as a function of the information content, from the information matrix, the equation (5), as:

$$E_k = Z(\omega_k)T \quad \underline{\phantom{xxx}} \quad (8)$$

Information couples to the frequency of a wave packet weighted by the relevant probability amplitude and may lead to either generation of information, radiation or matter, in the form of pairs of quarks or electrons, and the consequent processes, whether leading to lepton pairs, quark pairs or an entangled couple of information with an electromagnetic field quantum, can be depicted in the form of relevant Feynman diagrams, as given in Figure 1. In effect, the process and relevant diagrams are equivalent to the usual photon-led pair production mechanism.

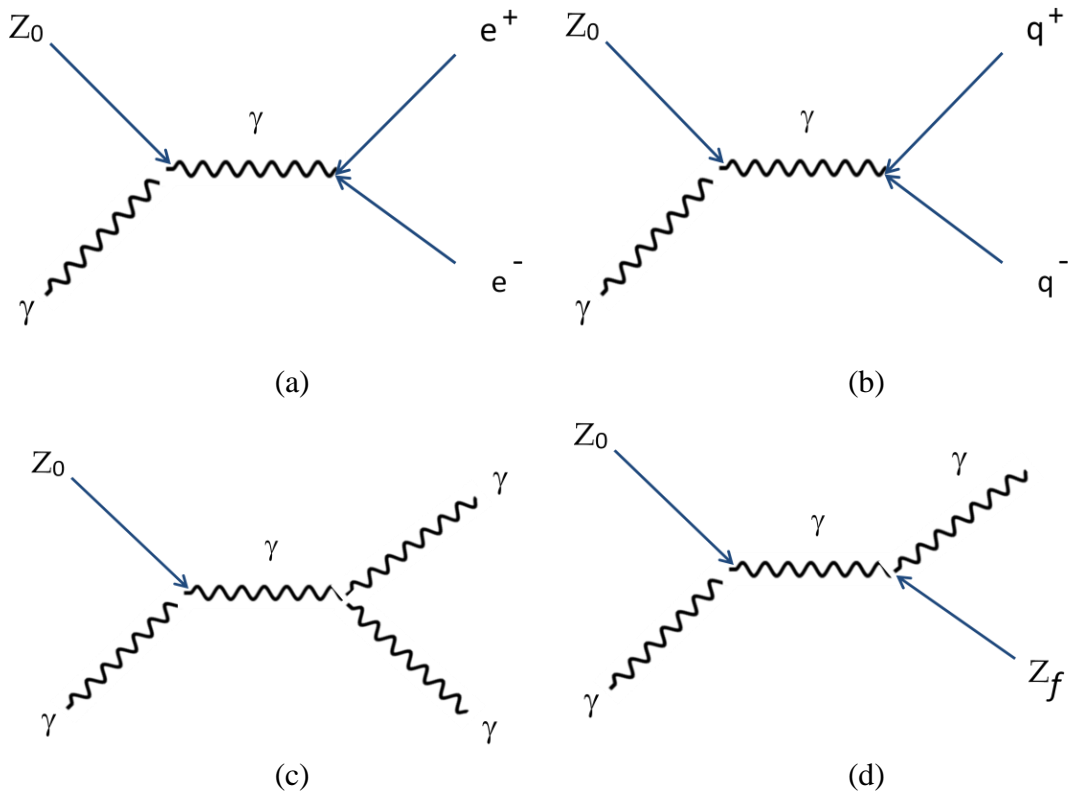

(a)     (b)     (c)     (d)

Figure 1. Four possible modes of interaction. Matrix elements ($\mathcal{M}_{fi}$) for these can be calculated as per the standard field theoretical methods.

In terms of Hamiltonian formulation of the complete quantum system, the field quanta and the information interacting with each other, if one were to identify a scalar field of information, a suitable Hamiltonian (with minimum terms, and assuming hermitian conjugates) could be written as:

$$\mathcal{H} = (a^\dagger a)\hbar\omega_k + \hbar\omega_k \zeta_n + \frac{1}{2}\lambda\zeta_n(a^\dagger + a)\hbar\omega \quad \underline{\quad\quad\quad} \quad (9)$$

The first term is the electromagnetic field, the second is the scalar information *"field"*, and the last term is the interaction term between the two. The $a^\dagger$ and $a$ are the usual creation and annihilation operators, which create individual field quanta or turn them back to the vacuum, respectively.

Addressing the measurement-oriented or experimental aspects of the problem, i.e. how to measure the signal arising out of the action of information, an immediate question would be the power spectrum. One would like to estimate a real-world physical signal corresponding to a word of information as a result of a measurement carried out over the quantum vacuum. Although information cannot be measured in its entirety or with exactness, since in a physical system it is entangled with the electromagnetic fields, and vice versa, if a measurement were to be performed with the aim to measure the information content in a signal, one can only measure an approximate value, which would have inherent in it a number of physical terms.

We thus formulate here an approximate expression for the power as a function of the information contained in a signal (beyond the heat current noise).

A time-evolving signal power is defined as:

$$\mathrm{H}(t) = \dot{Z}(\omega_k)T + MX(\omega) - \Gamma(t) \quad \underline{\quad\quad\quad} \quad (10)$$

Here $X(\omega)$ is noise spectral density of the heat current and M is its amplitude. $\Gamma(t)$ is the damping and back-action term which measures the loss of power to the environment.

The measured signal power (if a detector were to measure it) would be equal to:

$$\mathrm{H}(t) = \frac{Z(\omega_k)T}{t} + MX(\omega) - \Gamma(t) \quad \underline{\quad\quad\quad} \quad (11)$$

The form of noise spectral density can be obtained from the (frequency-dependent) noise of heat current, as given by any of the pertinent expressions [9], such as in a form:

$$X(\omega) = \frac{K}{\pi}\left[\frac{\hbar\omega}{2} + \frac{\hbar\omega}{\exp(\hbar\omega/kT)-1}\right] \quad \underline{\quad\quad\quad} \quad (12)$$

Here, K is the constant for the material properties of the channel and k is the Boltzmann constant. At zero temperature, the noise is restricted to intrinsic quantum fluctuations of the vacuum and satisfies the *Fluctuation-Dissipation Theorem*. However, the best description for this can only be obtained from statistical mechanics.

This expression gives an expected measured value, taking into account both deterministic and probabilistic contributions, as well as carries the contributions of the heat current (the ubiquitous noise which does not vanish at absolute zero).

**Conclusion and Discussion:**

It is extremely important to understand the physical basis and structure of information and its connection with quantum fields and other physical entities, including the thermodynamic underpinnings and implications of information giving rise to, and entangled with, the quantum fields. Geometry and probability both may play a pivotal role in these phenomena, and possibly inertia also, in the form of an emergent property (similar to time).

A century ago Shannon was concerned with the problem of recovering a useful piece of information (*data*) conveyed over a viable communication channel. A century later, the problem we encounter, is to understand what information is and how it relates to the origin and form of matter and radiation we have in the universe, especially in relation to the quantum field (or alternatively the String) formulation we have to describe matter and interactions. We also contend with the content of a legitimate signal beyond Johnson noise, and in addition to it, the problem of extracting the useful physical information (and not just the semantic content of it) out of a signal, which is measured in a laboratory in the form of power spectral density.

The information may itself be entropy or noise or they may be forms of it, however the information which translates into physical action, into the creation of bosonic or fermionic (matter) fields, must be of a characteristic nature with deep underlying symmetries, and distinguishable from incessant noise. Therefore, it may be possible (albeit quite difficult) to detect its presence in a signal, beyond all kinds of noise, with extremely sensitive measurement techniques. Such techniques are present and constantly being developed to detect extraordinary excitations within vacuum at very low temperatures, for instance and not limited to the detection of axions and very light cold dark matter particles [3]. A Fourier-transformed time series array of measured values from an isolated volume of vacuum at extremely low temperature (on the order of milli-Kelvin) with the help of a Josephson amplifier-based sensor or any other similar sensitive device capable of measuring minuscule powers would yield the presence of patterns of spectral information, distinct from thermal and systemic noise, upon careful analysis.

It was perhaps none other but Schrödinger himself who pondered upon and gave the first philosophical description of quantum entanglement [10]. The nature of reality cannot be completely understood or measured, the state of a physical system is entangled (a mixed state depending upon probability weights) between the information and field quanta, a pure quantum or information state, once risen out of a pure vacuum state, either does not exist or cannot be measured.

The virtual particles or pairs in vacuum, which have always been a mystery, are thus most likely nothing but the information and the source of it, they condense into photons and into the quark-anti-quark and electron-positron pairs under the well-understood pair production mechanisms, under the influence of electromagnetic interaction.

Somehow, following these arguments, it may be possible that under a physical mechanism, which is both deterministic and stochastic, information condenses into the fields of matter, radiation or both, and thus an information-condensation is possible which generates the energy and mass for all the fields within nature. A deterministic information code (based on geometry) as well as probabilities and chaos (especially owing to quantum fluctuations) within the quantum vacuum play their respective roles in this condensation. The fields generated in this way are thus fluctuating information condensates in their most fundamental form. However, we unfortunately are not aware of the underlying field generation symmetry groups and how it takes place.

The repercussions of physical information content of signals are extremely important. As it has been cited before, and the idea has been gathering momentum steadily, that everything has its base in information- the Wheeler's aphorism of *"It from Bit"*. This implies that the information is at the base of everything, including the quantum fields. Hence, identifying a signal in terms of its physical information basis is synonymous to understanding the origin and form of quantum fields, which was the main motivation of this model.

Based upon a simple model of elementary excitations within the quantum vacuum, we have endeavored to present a critique of meaningful words (or the suggested quanta of information) and elementary quantum fields. The model may also have important implications on our understanding of the initial events and the involved singularity which led to the creation of our universe (often dubbed as the *"Big bang"* under the current $\Lambda$-Cold Dark Matter model of physical cosmology), and its later evolution out of the primordial density fluctuations. The next step seems to understand the topological structure and common geometrical basis of physical information and field quanta.